# Perturbative Deformation Mechanism of Metasurface

Tuo Li[1,*], Xin Liu[1], Lin Zhu[1], and Lei Liang[1]

1. School of Electronic Engineering, Xi'an University of Posts and Telecommunications, Xi'an 710121 China

*Corresponding author: cyantl@163.com

ABSTRACT: Metasurfaces enable precise manipulation of light-matter interactions, and meta-atom coupling and scaling dominates their resonant properties and functional responses. Conventionally, although theories such as CMT and CDT can explain many metasurface phenomena, a unified mechanism that reveals how the deformation of metasurface affect its performance remains lacking. Here, by combining transformation optics and spatial perturbation, we proposed a universal deformation mechanism of metasurface, which establishes a general physical picture. Based on this mechanism, we interpret the resonance frequency shift caused by coupling of the meta-atoms, clarify the tuning law of resonant frequency via geometric scaling of unit structures, and further demonstrate the anisotropic shift of grating resonant peak. Theoretical predictions show consistency with full-wave simulation results in all three scenarios. Given the broad applicability of the transformation optics and perturbation theory, the universal mechanism with intuitive physical picture should be widely existed in diverse fields including photonics crystals, Bragg fibers, two-dimensional materials and crystalline optical properties.

**Key words:** Metasurface; Grating; Transformation optics; Coupling and Scaling;

Metasurfaces, artificially constructed arrays of subwavelength resonant meta-atoms, offer unparalleled capability to manipulate light–matter interactions, laying the groundwork for disruptive photonic technologies spanning diverse research domains [1–4]. Geometric deformation of individual meta-atoms also drastically alters the functional characteristics of metasurfaces. For example, spatial deformation in the interspacing of meta-atoms can induce coupling effects, which further modulate the overall electromagnetic performance of metasurfaces.

Now，two distinct theoretical paradigms have dominated the description of coupling in metasurface resonant arrays. The first is the mode energy coupling framework, which describes energy exchange between discrete modes via coupling coefficients, encompassing coupled-mode theory (CMT) [5-7], coupled dipole/multipole theory (CDT) [8-10]. CMT assumes coupling exists a priori and constrains its strength solely via energy conservation and time-reversal symmetry [11-13]. As a phenomenological approach, it can accurately fit observed frequency shifts but cannot provide a first-principles explanation for the fundamental origin of coupling [14-15]. CDT, by contrast, advances beyond CMT by seeking a microscopic mechanism for coupling: it approximates each meta-atom as a truncated set of electric and magnetic multipoles and calculates inter-element interactions via free-space Green's functions [16]. However, it remains a phenomenological theory: both the *ad hoc* truncation of higher-order multipoles and the a priori assumption that meta-atom responses are fundamentally multipolar are not rigorously derived from Maxwell's equations, leaving the first-principles origin of coupling unresolved. The second paradigm is periodic eigenmode analysis, comprising Bloch-mode theory and rigorous coupled-wave analysis (RCWA), which solve Maxwell's equations numerically under periodic boundary conditions [17]. While yielding highly accurate eigenfrequencies, these methods constitute numerical black boxes: they can only demonstrate that resonance shifts accompany changes in lattice constant, but provide no physical insight into the underlying mechanism or the fundamental nature of coupling itself.

In light of this, explaining the origin of coupling from the first principles of Maxwell's equations has become a widely pursued topic in the field [18-20]. Among these approaches, transformation optics offers an intuitive geometric framework for this challenge. Since both inter-atomic spacing-induced coupling and meta-atom deformation are fundamentally geometric modulations of the metasurface, transformation optics is inherently suited to handle such effects. However, existing transformation optics studies mainly employ analytical methods to treat these deformations rigorously [21–22]. For instance, conformal mapping yields exact analytical solutions [23], but it requires a rigorously analytical transformation and only limited set of geometries admit such solutions [24–25]. Hence, this approach is essentially confined to specific configurations. Here, we do not adopt the paradigm which requires rigorous analytical solutions by transformation optics. By combining the transformation optics and perturbative theory, a deformation perturbative mechanism with a unified picture is discovered, which provides an intuitive geometric perspective on the origin of meta-atom coupling. Moreover, beyond meta-atom coupling, this mechanism simultaneously reveals a geometric physical picture for understanding and predicting various seemingly distinct phenomena.

Figure 1 illustrates the meta-atom coupling mechanism via deformation perturbation. Representative unit cells (Figs. 1d and 1f) from metasurfaces A and B (Figs. 1a and 1b) have side lengths $p_1$ and $p_2$, respectively, with identical cylindrical meta-atoms of diameter $w_1$. The two configurations differ only in their gap dimensions.

The physical picture of meta-atom coupling revealed by the mechanism is shown in Fig. 1, which involves transforming metasurface B into B′ by stretching its gaps to match the period of metasurface A. Based on transformation optics, B and B′ are electromagnetically equivalent, allowing the study of the A−B relationship to be reformulated as the A−B′ relationship. Since A and B′ share identical periodicities, the only difference is the permittivity of the gap space ($\varepsilon_{eq}$) due to space stretching. For small variations, this permittivity change is treated as a perturbation. Thus, perturbation theory using metasurface A as a reference yields the resonant frequency of metasurface B′, which equals that of metasurface B due to their electromagnetic equivalence. As illustrated in Fig.1 the mechanism reveals a new geometric perturbation perspective that the coupling arises from a perturbation effect induced by spatial deformation.

We provide a quantitative description of the coupling by using perturbative deformation mechanism. In a coordinate system with the metasurface in the $x$-$y$ plane and normal incidence along $z$ (Fig. 1), a unit cell of period $p_1$ contains a dielectric cylinder (radius $w_1$) and an air gap of area $S_p = p_1^2 - \pi w_1^2$ (Fig. 1d). Mapping from Fig. 2(f) to Fig. 2(e) stretches only the air gap region, resulting in a change in the dielectric properties of the air gap. According to the transformation optics, the dielectric tensor can be calculated by $\varepsilon = J\varepsilon_0 J^T / \det(J)$. Under perturbation, it can be demonstrated that the dielectric tensor reduces to a scalar. From this scalar, $\varepsilon_{eq}$ can be further derived (Detailed analysis of the scalar and derivation of Eq. (1) are provided in the Supplement 1).

$$\varepsilon_{eq} = \varepsilon_0 \frac{S_v}{S_p} = \varepsilon_0 \frac{p_2^{\ 2} - S_{\text{unit}}}{p_1^{\ 2} - S_{unit}} \tag{1}$$

Here, $S_{\text{unit}}$ represents the area of the meta-atom (i.e., in Fig. 1, $S_{\text{unit}}=\pi w_1^2$), and $p_1$ and $p_2$ indicate the period of the metasurface in Fig.1(e) and Fig.1(f).

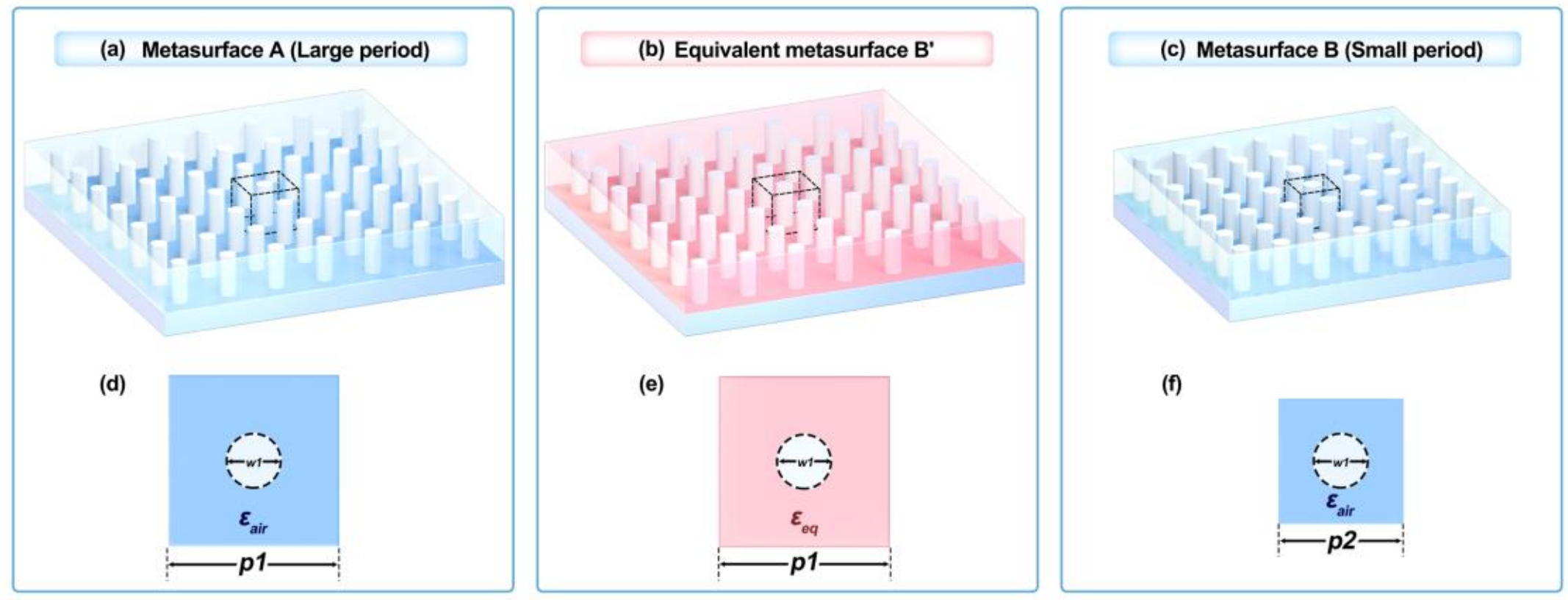

**Fig. 1.** Physical picture of meta-atom coupling revealed by the mechanism: (a)-(c) shows three metasurfaces; (d)-(e) are the unit cells extracted from the three metasurfaces.

Fig. 1(e) can be regarded as a slight permittivity perturbation of the air gap in Fig. 1(d). Eq. (1) has already yielded the permittivity $\varepsilon_{eq}$ in Fig.1(e). The gap in Fig. 1(d) is filled with air, whose relative permittivity is approximately equal to 1 ($\varepsilon_{r,air}\approx 1$). Hence, the change in the relative dielectric constant is $\Delta\varepsilon_r = \varepsilon_{eq}/\varepsilon_0 - 1 = \left(p_2^{\,2} - p_1^{\,2}\right)/\left(p_1^{\,2} - S_{unit}\right)$ . It can be seen that $\Delta\varepsilon_r$ is a constant. According to the perturbation theory of the Rayleigh quotient [26], it can be obtained:

$$\frac{\Delta f}{f_0} = -\frac{\varepsilon_0}{2W}\int_{\Delta V_{gap}} \Delta\varepsilon_r \left|\vec{E}_0\right|^2 dV \approx c_{\text{gap}}\left[1-\beta^2\right] \tag{2}$$

Here, $\beta=p_2/p_1$ is the scaling factor, which serves as the scaling factor of unit cell of the metasurface. $\beta<1$ indicates period compression and $\beta>1$ indicates period stretching. $f_0$ represents the resonant frequency of the metasurface $A$ with a period of $p_1$. $f$ represents the resonant frequency of the metasurface B′ with a period of $p_2$. $\Delta f=f-f_0$ represents the change in resonant frequency due to variations in coupling strength. Let $c_{gap} = \varepsilon_0 \iint_{\Delta Vgap} \left|\vec{E}_0\right|^2 dV \Big/ 2p_1^{\,2}\left(p_1^{\,2} - S_{unit}\right)W$ . $W$ denotes the total energy distributed over one period of the metasurface. When the change of the gap space involved in the above transformation is sufficiently small to be treated as a perturbation, the ratio of the energy stored in the air gap to the total energy $W$ within one period of the metasurface is approximately constant. Hence, $c_{gap}$ is a constant. Moreover, since $p_1^{\,2} - S_{unit} > 0$ , $c_{gap}$ is a positive constant. By obtaining the resonant frequencies of two metasurfaces with different periods via full-wave simulation, $c_{gap}$ can be solved from Eq. (2).

To verify the validity of the above theoretical analysis, we conducted numerical calculations using the full-wave electromagnetic simulation software CST Studio Suite 2025 and are compared with the theoretical results. The coordinate system of the full-wave electromagnetic simulation process was consistent with the theoretical part, that is, the electromagnetic field was incident perpendicularly, the propagation direction of the electromagnetic field was the $z$-axis, and the metasurface was in the $x$-$o$-$y$ plane.

Three metasurfaces with square (2 μm × 2 μm × 3 μm, Si), cylindrical (r = 1 μm, h = 3 μm), and hexahedral (a = 1 μm, h = 3 μm) meta-atoms were tested. Figure 2 shows that meta-atom sizes are fixed while resonance peaks shift via period adjustment. Varying the period compresses the air gap, altering inter-unit coupling and shifting the peaks. Figures 2(a)–(c) show the resonance spectra of three metasurfaces with square, cylindrical, and hexahedral meta-atoms, respectively. As the period decreases, all three exhibit significant blue shifts, consistent with Eq. (2). Figures 2(d)–(f) compare theory (curves) with CST simulation results (points) across $\beta \in (0.9, 1.1)$, showing good agreement and validating the perturbative deformation mechanism in meta-atom coupling. The above results demonstrate the validity of the perturbative deformation mechanism in meta-atom coupling.

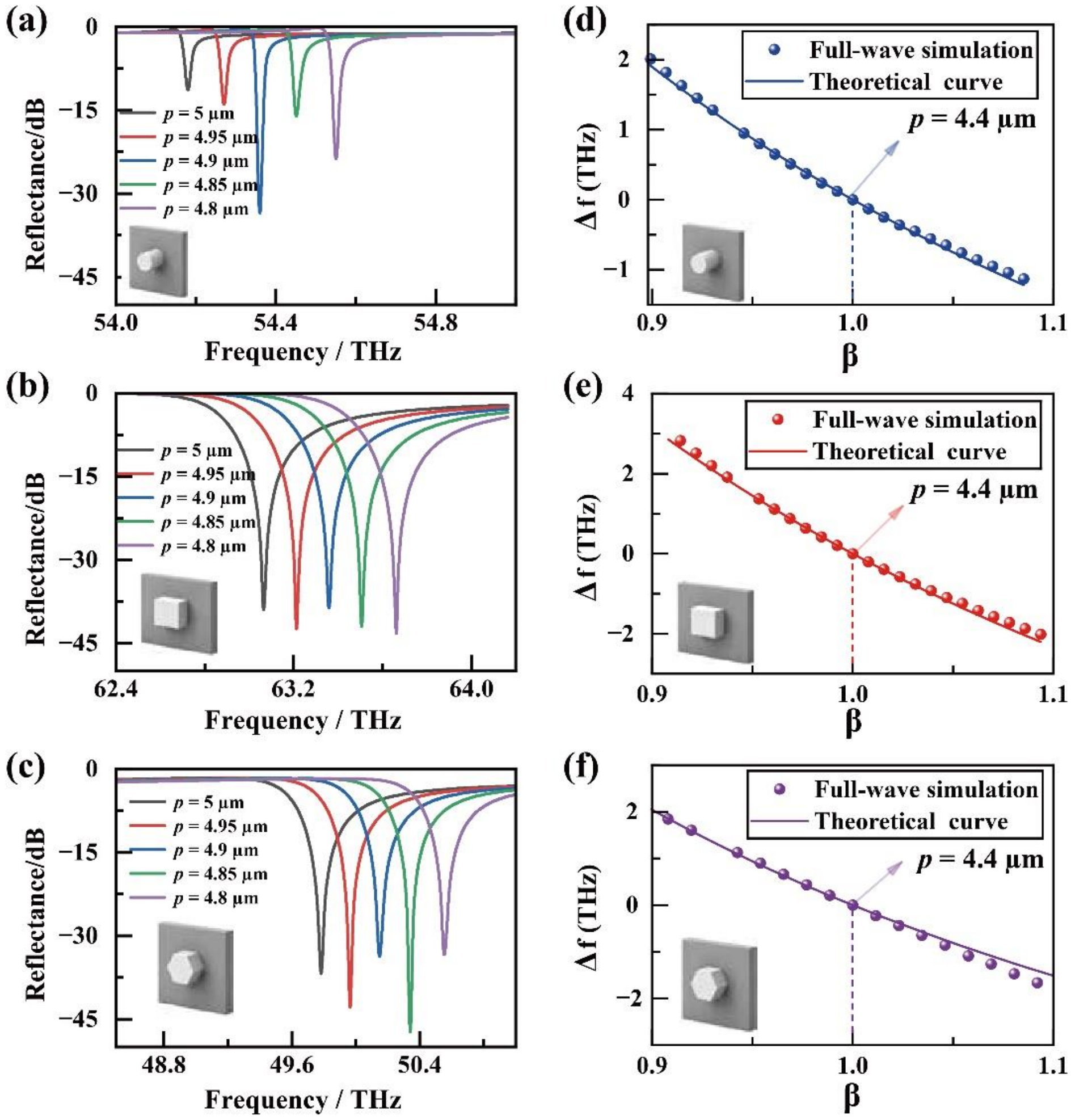

**Fig. 2.** Meta-atom dimensions are fixed, while the resonant peak is tuned via the metasurface period. (a)–(c) show spectra of three metasurfaces with different meta-atom structures; (d)–(f) compare theoretical results with full-wave simulations.

Moreover, this mechanism extends beyond metasurface coupling to grating coupling and explains resonance shifts under varying polarization (Appendix A). Figs. 1–2 show 2D gap deformation case. The perturbative mechanism is not limited to this case and offers richer degrees of freedom (Supplementary Materials 2).

Next, the perturbative deformation mechanism is further used to unify the phenomenon of meta-atom scaling at a fixed metasurface period. Figure 3 illustrates the physical picture of the phenomenon revealed by the mechanism. Figs.3(d)-(f) are one unit cell extracted from Figs. 3(a)-(c). The period is fixed at $p_1$, with unit diameters $w_1$ (Fig. 3(d)) and $w_2$ (Fig. 3(f)) for metasurfaces A and B, respectively. The radius difference makes direct correlation between metasufrace A and B nontrivial. This mechanism resolves this by transforming metasurface B′ to match the unit cell of metasurface A. Via transformation optics,

B′ is electromagnetically equivalent to B, reducing the problem to comparing A with B′. As seen in Figs. 3(d)–(e), the only distinction is a slight permittivity variation in the meta-atom. Perturbation theory can then quantify its impact. Although Figures 1 and 3 depict seemingly distinct phenomena, they share identical physical processes under the deformation perturbation mechanism.

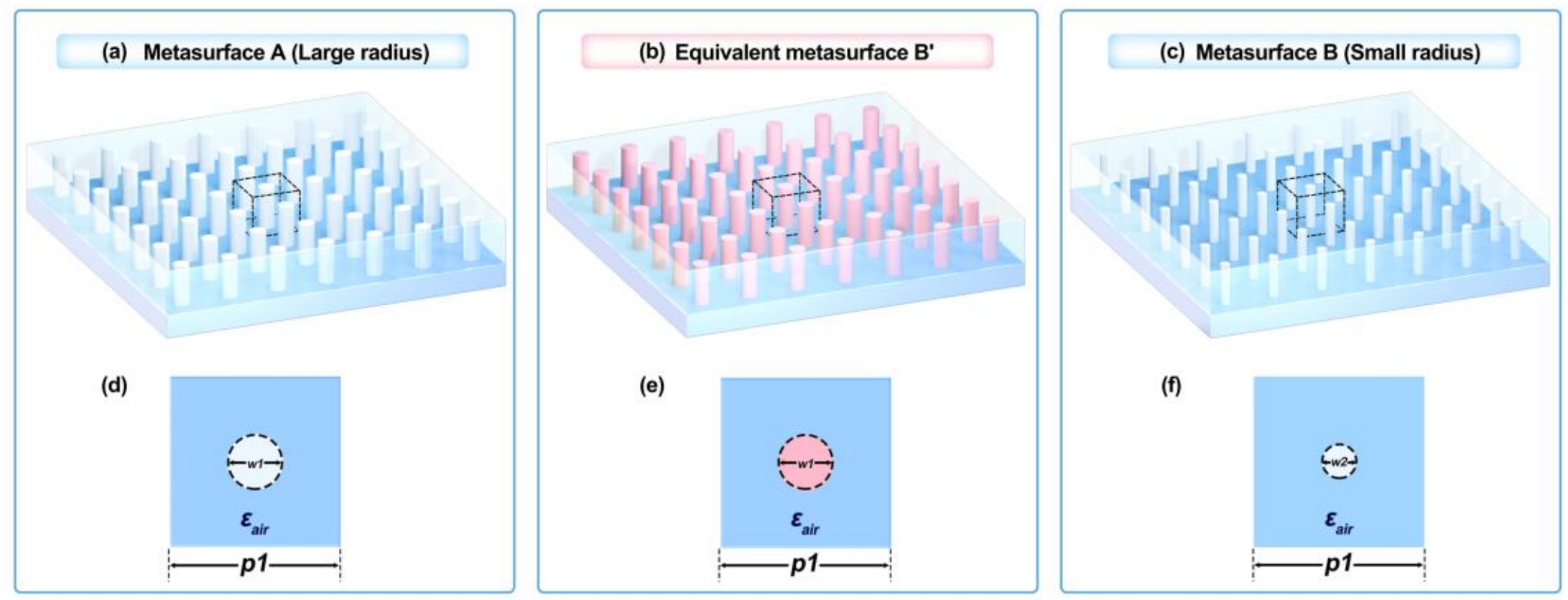

**Fig. 3.** Physical picture of meta-atom scaling revealed by the mechanism: (a)-(c) shows three metasurfaces; (d)-(e) are the unit cells extracted from the three metasurfaces.

The following provides a quantitative description of meta-atom scaling via the theory. Analogous to the proof of Eq. (2), this scaling (i.e., the spatial deformation from B to B′) constitutes an approximate conformal transformation under perturbation. The sole source of error arises from neglecting air gap deformation. However, since the scaling is inherently perturbative and the air gap dimensions greatly exceed the meta-atom size, this effect is safely neglected. A similar derivation yields Eq. (3):

$$\frac{\Delta f}{f_0} = -\frac{\varepsilon_0}{2W}\int_{\Delta V unit} \Delta\varepsilon_r \left|\vec{E}_0\right|^2 dV \approx c_{unit}\left[1-\beta^2\right] \tag{3}$$

where $\beta = w_2/w_1$ denotes the scaling factor, with $w_2$ and $w_1$ representing the diameters of metasurface A and metasurface B, respectively (shown in Figs. 3(d)–(f)). $f_0$ represents the resonant frequency of the metasurface $A$ with a period of $w_1$. $f$ represents the resonant frequency of the metasurface $B'$ with a period of $w_2$. $\Delta f=f-f_0$ represents the change in resonant frequency. Let $c = \varepsilon_0\varepsilon_r \iint_{\Delta V unit} \left|\vec{E}_0\right|^2 dV \Big/ 2W$ . Analogous to $c_{gap}$ in Eq. (2), $c_{unit}$ can be approximated as a constant under perturbative conditions.

Figure 4 illustrate the period of metasurface remains unchanged, and resonance peaks shift by adjusting the size of meta-atom. Figures 4(a)–4(f) show three silicon metasurface units: square pillar (2 μm × 3 μm), cylinder (r = 1 μm, h = 3 μm), and hexahedron (edge = 1 μm, h = 3 μm). As the meta-atom size decreases, the resonance peaks of all three metasurfaces shift significantly toward the blue, consistent with Eq. (3). To further verify Eq. (3), CST parameter scans were performed over a scaling factor $\beta \in (0.9, 1.1)$. Figures 4(d)–4(f) compare the full-wave simulation (data points) with theoretical predictions (curves), showing excellent agreement. The above results further demonstrate the validity of the mechanism to describe influence of meta-atom dimension variation on resonant frequency shift.

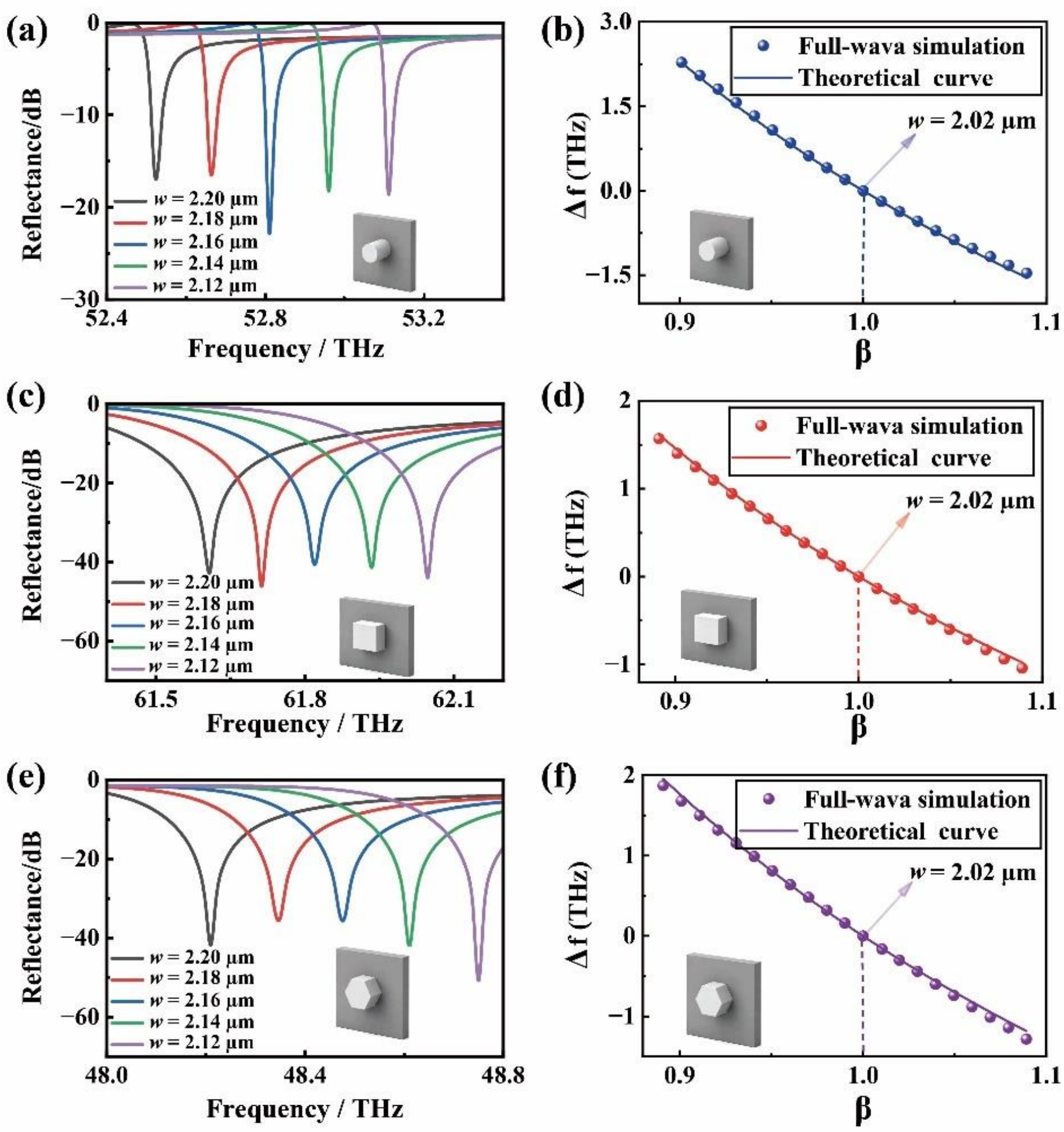

**Fig.4.** Period of metasurface remains unchanged, and resonance peaks shift by adjusting the size of meta-atom. (a)-(c) show resonance spectra of three metasurfaces with distinct meta-atoms. (d)–(f) present a comparison between the theoretical results and the full-wave simulation results.

The aforementioned perturbative deformations—arising from meta-atom coupling, meta-atom scaling and grating geometry—can all be described within a unified framework. Notably, this framework extends beyond these three cases (see Supplement materials 2-3). Resonant frequency tuning in metasurfaces can be further enriched by modulating meta-atom height, side length, cross-section, volume, or inter-element air gaps. Therefore, this work offers a general physical picture for perturbative deformation of metasurface. Although a general picture is revealed, it is worth noting that theory of the mechanism is based on perturbation theory, so it has a certain applicable range. The mechanism may become inapplicable once exceeding the perturbation limit. Additionally, the perturbation calculation in the mechanism employed here is not limited only to the Rayleigh-quotient-based perturbation theory in Eq. (2) and Eq. (3). It may be carried out from the perspective of Maxwell eigenmodes as discussed in Refs. [18–20], potentially enabling simultaneous treatment of multiple modes and even lossy systems.

In summary, we have discovered a perturbative deformation mechanism that reveals a unified picture for metasurface geometrical deformation. Leveraging this, richer resonant control is achieved via deliberate spatial deformations (gratings, air gaps, meta-atom), with theoretical predictions showing excellent agreement with full-wave simulation results. The description of the mechanism is based on transform optics and perturbation theory. Transformation optics is derived from the covariance of Maxwell's equations without any approximation and has a wide application scope. Perturbation theory is a mathematical theory with universal applicability as well. Therefore, the aforementioned mechanism similarly has broad applicability and holds promise for extension to photonic crystals, Bragg fibers, two-dimensional materials, and crystalline optical properties.

Acknowledgements—The authors wish to acknowledge Prof. Jun Dong (Xi'an University of Posts and Telecommunications) for his meaningful advice on our research. This work was supported by the National Natural Science Foundation of China (Grants No. 12304329), in part by Qualification Number of Shaanxi Provincial Key Technology Research (Grants No. 2025CG-GJHX-02), Natural Science Foundation of Shaanxi Provincial Department of Education (Grants No. 23JK0667).

## Appendix: Origin of Grating Resonance Anisotropy Revealed by the Mechanism

Grating is an important element in optics and can be equivalently modeled as a 1D metasurface. We test the mechanism on gratings. As shown in Fig. A1, varying the period changes the gap between dielectric bars, thus tuning the resonant frequency at fixed bar dimensions. The resonance also responds anisotropically to polarization, with different directions exciting distinct modes. A Cartesian coordinate system is defined as follows: the longitudinal extension of dielectric bars aligns with the *x*-axis, the crosswise direction orthogonal to the bars corresponds to the *y*-axis, and electromagnetic wave propagation runs along the *z*-axis. For normal incidence of electromagnetic waves along the *z*-axis, two canonical excitation modes arise based on electric field polarization: one with the electric field parallel to the dielectric bars (*x*-polarization), and the other with the electric field perpendicular to the bars (*y*-polarization). Driven by the intrinsic anisotropy of the grating structure, these two polarization inputs independently excite resonant eigenmodes at their respective characteristic resonant frequencies.

We quantitatively analyze how the gap deformation between dielectric bars influences anisotropic shift of grating resonant peaks. Since the grating analysis is relatively simple, we do not elaborate on its physical picture here. Instead, an intuitive physical picture is presented in in Figure 1 and Figure 3. The mechanism is discussed below. Let the initial period of the grating be $\Lambda_0$ and the tuned period be $\Lambda$, with a period scaling factor $\beta=\Lambda/\Lambda_0$ introduced. The scaling process keeps the width of the dielectric bars (hereafter referred to as meta-atoms) unchanged. For $\beta<1$, the inter-cell spacing is compressed, whereas for $\beta>1$, it is stretched. A linear coordinate transformation is applied to the gap of the meta-atom. The Coordinate transformation from the virtual space to the physical space： $x = x, y = y'/\beta, z = z'$ , where ($x$, $y$, $z$) denotes the physical space, $(x', y', z')$ denotes the virtual space. The virtual space is air which the permittivity $\varepsilon'$ is approximately equal to $\varepsilon_o$. $J = \partial(x, y, z)/\partial(x', y', z')$ represent the Jacobian matrix. Based on the relationship of coordinate mapping, it can be derived that the Jacobian matrix is equal to $J = diag(1, 1/\beta, 1)$ . Since $J$ is a diagonal matrix and thus its determinant is equal to $\det(J)=1/\beta$. According to transformation optics, the expression for the dielectric tensor in physical space is given by:

$$\varepsilon = \frac{J\varepsilon' J^T}{\det(J)} = \varepsilon_0 \begin{bmatrix} \beta & 0 & 0 \\ 0 & 1/\beta & 0 \\ 0 & 0 & \beta \end{bmatrix} \tag{A1}$$

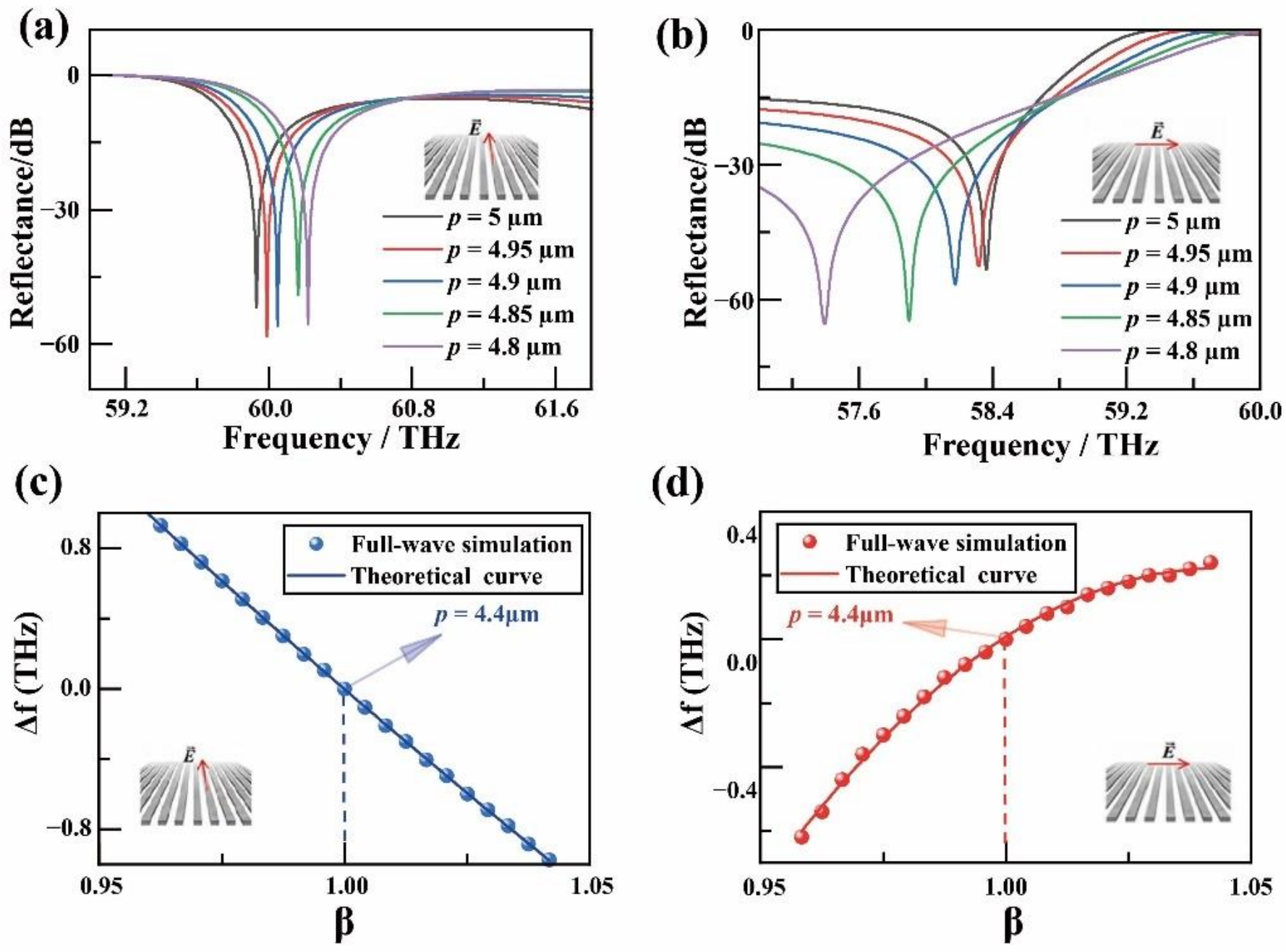

Fig. A1. Polarization-dependent frequency shifts of the resonant frequency in grating. (a) and (b) show the resonant frequency curves under *x*- and *y*-polarized incidences at different grating periods, respectively. (c) and (d) are the theoretical curve and the full-wave simulation result for *x*- and *y*-polarized incidences under different scaling factors, respectively.

As we known, when an *x*-polarized electromagnetic wave is normally incident on the grating surface, only $E_x$ is generated inside the grating, so polarization occurs solely along the *x* direction. In this case, the permittivity corresponding to the *x*-direction polarization response is $\varepsilon_0\beta$. When a *y*-polarized electromagnetic wave is incident, both $E_y$ and $E_z$ are simultaneously generated inside the grating, so polarization occurs along both the *y* and *z* directions. In this case, the permittivity's corresponding to the *y*- and *z*-direction polarization responses are $\varepsilon_0/\beta$ and $\varepsilon_0\beta$, respectively. Thus, the variation of the relative permittivity can be expressed as:

$$\Delta\varepsilon_r = \begin{cases} diag\left(\beta-1,0,0\right), x-polarization \\ diag\left(0,1/\beta-1,\beta-1\right), y-polarizaiton \end{cases} \tag{A2}$$

When the scaling factor $\beta$ is fixed, $\Delta\varepsilon_r$ is a constant. According to the perturbation theory of the Rayleigh quotient, Eq. (A3) can be obtained:

$$\frac{\Delta f}{f_0} = -\frac{\varepsilon_0}{2W}\int_{Vgap}\Delta\varepsilon_r\left|\vec{E}_0\right|^2 dV \approx \begin{cases} c_x\left(1-\beta\right), x-polarization \\ c_y\left(1-1/\beta\right)+c_z\left(1-\beta\right), y-polarizaiton \end{cases} \tag{A3}$$

where $f_0$ and $f$ denote the resonant frequencies at grating periods of $\Lambda_0$ and $\Lambda$, respectively. $\Delta f=f-f_0$ denotes the shift in resonant frequency after the gap is distorted. $V$ is the volume of gap, $\beta=\Lambda/\Lambda_0$ is the scaling factor. $W$ denotes the total electromagnetic energy distributed over one period of the grating. When the change of the gap space involved in the above transformation is sufficiently small to be treated as a perturbation, the ratio of the energy stored in the air gap to the total energy $W$ within one period of the grating is approximately constant. Thus, the $c_i = \varepsilon_0\iint_{\Delta Vgap}\left|E_{0i}\right|^2 dV \Big/ 2W\,(i=x,y,z)$ can be seen constants. Note that $c_i$ an unknown constant. By obtaining the resonant frequencies of three metasurfaces with different periods via full-wave simulation, $c_i$ can be solved from Eq. (A3).

Full-wave simulations in CST Studio Suite 2025 were conducted to validate the theoretical analysis by comparing simulation results with theoretical predictions. The geometric parameters of the grating are set as follows: silicon strip width of 1 μm, thickness of 1 μm, and permittivity of 12. The initial period of the grating is 5 μm (corresponding to an air gap of 4 μm between adjacent silicon strips), and the period is continuously tuned by adjusting the air gap width. Figure A1(a) presents the resonant frequency curves under $x$-polarized incidence at different period of the grating. The silicon strip remains unchanged, only the air gap is changed. It can be observed that as the period decreases (i.e., the air gap is compressed), the resonant peaks exhibit a pronounced blue shift, which is in good agreement with the prediction of Eq. (A3). To further validate Eq. (A3), a parametric sweep study was conducted. Figures A1(c) and A1(d) present the full-wave simulation results and theoretical predictions of the resonant frequency under $x$- and $y$-polarized incidences, respectively, where the data points represent the results obtained from the full-wave simulation and the curves denote the theoretical predictions. As shown in Fig.A1(c) and Fig.A1(d), when the value of β is in the region of (0.95, 1.05), the theoretical predictions are in consistent with the full-wave simulation results, thereby confirming the physical mechanism of meta-atom coupling—namely, that spatial deformation induces perturbation. This mechanism explains the polarization-dependent blue-red shifts in resonant frequency, confirming the structure's anisotropy.

**Reference**

[1] A. H. Dorrah, N. A. Rubin, A. Zaidi, M. Tamagnone, and F. Capasso, Metasurface optics for on-demand polarization transformations along the optical path, Nat. Photon. 15, 287 (2021).

[2] S. Q. Li, C. Chen, G. X. Wang, S. Y. Ge, J. Q. Zhao, X. S. Ming, W. Zhao, T. Li, and W. F. Zhang, Metasurface polarization optics: phase manipulation for arbitrary polarization conversion condition, Phys. Rev. Lett. 134, 023803 (2025).

[3] E. M. Renzi, E. Galiffi, X. Ni, and A. Alù, "Hyperbolic Shear Metasurfaces," Phys. Rev. Lett. 132, 263803 (2024).

[4] M. Xu, Q. He, M. Pu, F. Zhang, L. Li, D. Sang, and X. Luo, Emerging long-range order from a freeform disordered metasurface, Adv. Mater. **34**, 2108709 (2022).

[5] H. A. Haus, Waves and Fields in Optoelectronics (Prentice-Hall, Englewood Cliffs, NJ, 1984).

[6] H. A. Haus and W. Huang, Coupled-mode theory, Proc. IEEE 79, 1505 (1991). DOI：10.1109/5.104225

[7] W. Suh, Z. Wang, and S. H. Fan, Temporal coupled-mode theory and the presence of non-orthogonal modes in lossless multimode cavities, IEEE J. Quantum Electron. 40, 1511 (2004). DOI：10.1109/JQE.2004.834562

[8] A. B. Evlyukhin, C. Reinhardt, A. Seidel, B. S. Luk'yanchuk, and B. N. Chichkov, Optical response features of Si-nanoparticle arrays, Phys. Rev. B 82 , 045404 (2010). DOI：10.1103/PhysRevB.82.045404

[9] B. T. Draine and P. J. Flatau, Discrete-dipole approximation for scattering calculations, J. Opt. Soc. Am. A 11, 1491 (1994).

[10] M. A. Yurkin and A. G. Hoekstra, The discrete dipole approximation: An overview and recent developments, J. Quant. Spectrosc. Radiat. Transfer 106, 558 (2007). DOI：10.1016/j.jqsrt.2007.01.034

[11] S. G. Johnson, M. Ibanescu, M. A. Skorobogatiy, O. Weisberg, T. D. Engeness, M. Soljačić, S. A. Jacobs, F. Joannopoulos, and Y. Fink, Perturbation theory for Maxwell's equations with shifting material boundaries, Phys. Rev. E 66, 066608 (2002).

[12] S. Fan, W. Suh, and J. D. Joannopoulos, Temporal coupled-mode theory for the Fano resonance in optical resonators, J. Opt. Soc. Am. A 20, 569 (2003).
[13] Z. Fan, M. R. Shcherbakov, M. Allen, J. Allen, and G. Shvets, Perfect diffraction with multiresonant bianisotropic metagratings, ACS Photonics 5, 4303 (2018).
[14] I. Sersic, M. A. van de Haar, F. B. Arango, and A. F. Koenderink, Transformation optics for plasmonic nanoparticles beyond the quasistatic limit, Phys. Rev. Lett. 108, 223903 (2012). DOI: 10.1103/PhysRevLett.108.223903
[15] E. M. Purcell and C. R. Pennypacker, Scattering and absorption of light by nonspherical dielectric grains, Astrophys. J. 186, 705 (1973).
[16] D. R. Abujetas, J. Olmos-Trigo, J. J. Sáenz, and J. A. Sánchez-Gil, Coupled electric and magnetic dipole formulation for planar arrays of particles: Resonances and bound states in the continuum for all-dielectric metasurfaces, Phys. Rev. B 102, 125411 (2020).
[17] A. O. Sakin, H. Kurt, and M. Unlu, Exceptional-point engineered all-dielectric metasurface with tailored asymmetric scattering, Opt. Lett. 50, 610 (2025).
[18] F. Alpeggiani, N. Parappurath, E. Verhagen, and L. Kuipers, Quasinormal-mode expansion of Maxwell's equations: Green's functions and light–matter interactions in open nanophotonic systems, Phys. Rev. X 7, 021035 (2017).
[19] C. Sauvan, J. P. Hugonin, I. S. Maksymov, and P. Lalanne, Theory of the spontaneous optical emission of nanosize photonic and plasmon resonators, Phys. Rev. Lett. 110, 237401 (2013).
[20] J. Lin, M. Qiu, X. Zhang, H. Guo, Q. Cai, S. Xiao, Q. He, and L. Zhou, Tailoring the lineshapes of coupled plasmonic systems based on a theory derived from first principles, Light: Sci. Appl. 9, 158 (2020).
[21] J. B. Pendry, A. Aubry, D. R. Smith, and S. A. Maier, Transformation optics and subwavelength control of light, Science 337, 549 (2012).
[22] Zhang, X. Zhang, Y. Duan, L. Zhang, and X. Ni, "All-optical geometric image transformations enabled by ultrathin metasurfaces," Nat. Commun. 14, 8374 (2023)
[23] F. Yang, P. A. Huidobro, and J. B. Pendry, Transformation optics approach to singular metasurfaces, Phys. Rev. B 98, 125409 (2018).
[24] A. Aubry, D. Y. Lei, S. A. Maier, and J. B. Pendry, Interaction between plasmonic nanoparticles revisited with transformation optics, Phys. Rev. Lett.105, 233901 (2010). DOI: 10.1103/PhysRevLett.105.233901
[25] M. Kraft, Y. Luo, S. A. Maier, and J. B. Pendry, Designing Plasmonic Gratings with Transformation Optics, Phys. Rev. X 5, 031029 (2015)
[26] D. M. Pozar, Microwave Engineering, 4th ed. (Wiley, Hoboken, NJ, 2012).

# Supplementary Materials for Perturbative Deformation Mechanism of Metasurface

Tuo Li[1,*], Xin Liu[1], Lin Zhu[1], and Lei Liang[1]

1. School of Electronic Engineering, Xi'an University of Posts and Telecommunications, Xi'an 710121 China

*Corresponding author: cyantl@163.com

# S1: Derivation of Eq. (1)

The derivation of Eq. (1) requires the following three steps: First, we prove that under a small deformation, the transformation in Fig. S1 is approximately conformal transformation. Then, we derive the dielectric tensor under a conformal mapping. Further, we show that the longitudinal field component inside the metasurface is non-zero. Based on the above three steps, we obtain Eq. (1).

**First,** we prove that under the perturbation regime, the transformation on the shaded region in Fig. S1 is approximately a conformal transformation.

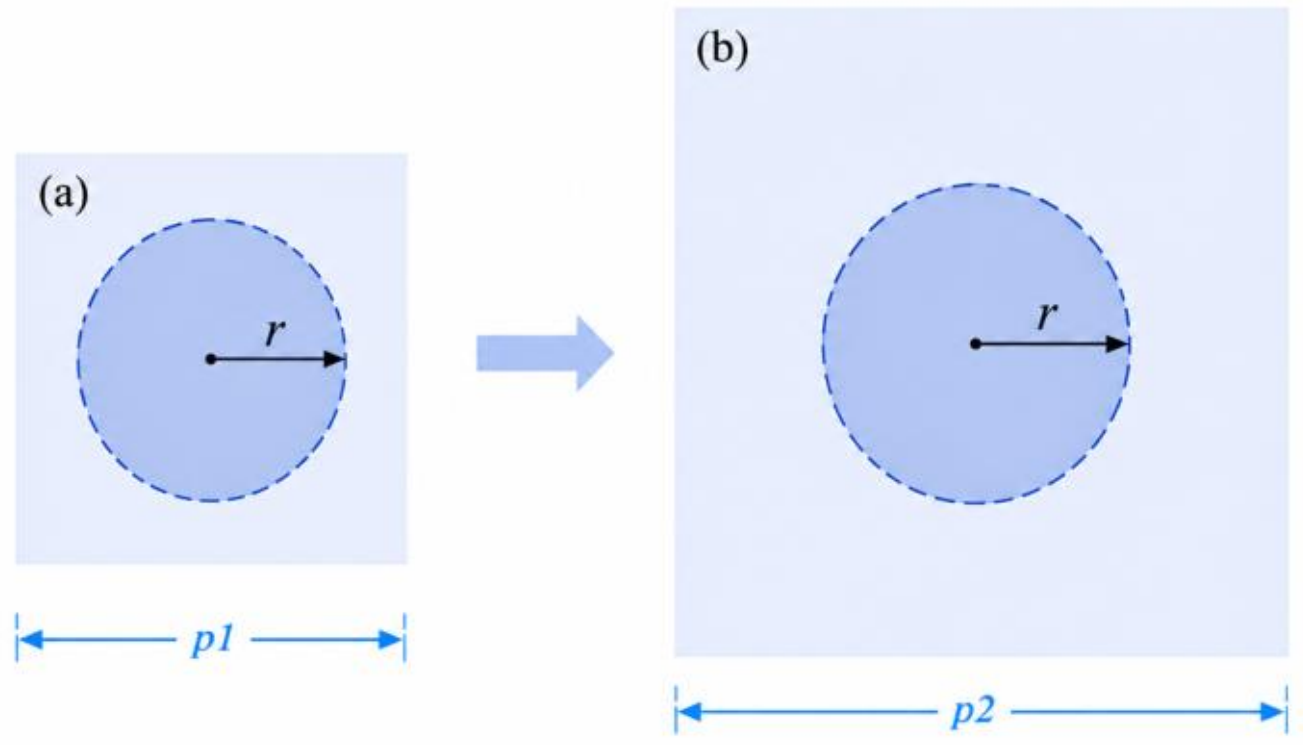

Fig. S1. Transformation of a unit cell of metasurface. (a) is the virtual space. (b) is physical space. Above transformation remains invariant only within the unit cell (blue region), while stretching is applied solely to the unit cell gaps (shaded region).

Consider the mapping $z = f(z')$ on the complex plane, where $z = x + iy$ denotes the physical space and $z' = x' + iy'$ the virtual space. This mapping features continuous first partial derivatives and nowhere vanishing derivative; hence, it is conformal provided the Cauchy–Riemann equations hold:

$$\frac{\partial x}{\partial x'} = \frac{\partial y}{\partial y'}, \frac{\partial x}{\partial y'} = -\frac{\partial y}{\partial x'} \tag{S1}$$

The physical significance of a conformal transformation lies in the fact that the stretching factors along the $x$ and $y$ directions are identical at every point, thereby preserving the angle between any two intersecting curves, which corresponds to an isotropic scalar medium. As shown in Fig. S1, the inner boundary radius $r_0$ remains fixed, while the outer boundary undergoes a radial linear scaling by a factor of $\varepsilon$. In polar coordinates, this transformation reads:

$$r(\theta) = r_0 + k(\theta)\cdot(r' - r) \tag{S2}$$

where $k(\theta)$ denotes the stretching coefficient:

$$k(\theta) = \frac{(1+\varepsilon)R'(\theta) - r_0}{R'(\theta) - r_0} \tag{S3}$$

where $\varepsilon$ denotes the relative stretching of the outer boundary (a 10% stretch corresponds to $\varepsilon$=0.1), and $R'(\theta)$ represents the radial length of the outer boundary in the virtual space (the outer boundary of the square is $R'(\theta) = p_1/2\cos\theta$, and $p_1$ is the side length of the square in the virtual space). The angular coordinate $\theta$ remains invariant in both spaces (The radial transformation does not alter the angle). Via the conversion between polar and Cartesian coordinates, the physical space is $x = r(\theta)\cos\theta, y = r(\theta)\sin\theta$ , and the virtual space is $r' = \sqrt{x'^2 + y'^2}$ , $\theta = \arctan(y'/x')$ . Taking the partial derivatives of the physical space coordinates with respect to the virtual space coordinates yields:

$$
\begin{aligned}
\frac{\partial x}{\partial x'} &= \frac{\partial x}{\partial r}\frac{\partial r}{\partial r'}\frac{\partial r'}{\partial x'} + \frac{\partial x}{\partial r}\frac{\partial r}{\partial \theta}\frac{\partial \theta}{\partial x'} + \frac{\partial x}{\partial \theta}\frac{\partial \theta}{\partial x'} \\
&= \cos\theta \cdot k(\theta) + \cos\theta \cdot (r' - r_0)\frac{dk}{d\theta} \cdot \left(-\frac{\sin\theta}{r'}\right) + (-r\sin\theta) \cdot \left(-\frac{\sin\theta}{r'}\right) \\
&= k(\theta)\cos^2\theta - \frac{(r' - r_0)\cos\theta\sin\theta}{r'} \cdot \frac{dk}{d\theta} + \frac{r\sin^2\theta}{r'}
\end{aligned}
\tag{S4}
$$

$$
\begin{aligned}
\frac{\partial y}{\partial y'} &= \frac{\partial y}{\partial r}\frac{\partial r}{\partial r'}\frac{\partial r'}{\partial y'} + \frac{\partial y}{\partial r}\frac{\partial r}{\partial \theta}\frac{\partial \theta}{\partial y'} + \frac{\partial y}{\partial \theta}\frac{\partial \theta}{\partial y'} \\
&= \sin\theta \cdot k(\theta) + \sin\theta \cdot (r' - r_0)\frac{dk}{d\theta} \cdot \left(-\frac{\cos\theta}{r'}\right) + (-r\cos\theta) \cdot \left(-\frac{\cos\theta}{r'}\right) \\
&= k(\theta)\sin^2\theta - \frac{(r' - r_0)\cos\theta\sin\theta}{r'} \cdot \frac{dk}{d\theta} + \frac{r\cos^2\theta}{r'}
\end{aligned}
\tag{S5}
$$

$$
\begin{aligned}
\frac{\partial x}{\partial y'} &= \frac{\partial x}{\partial r}\frac{\partial r}{\partial r'}\frac{\partial r'}{\partial y'} + \frac{\partial x}{\partial r}\frac{\partial r}{\partial \theta}\frac{\partial \theta}{\partial y'} + \frac{\partial x}{\partial \theta}\frac{\partial \theta}{\partial y'} \\
&= \cos\theta \cdot k(\theta) \cdot \sin\theta + \cos\theta \cdot (r' - r_0)\frac{dk}{d\theta} \cdot \left(-\frac{\cos\theta}{r'}\right) + (-r\sin\theta) \cdot \left(-\frac{\cos\theta}{r'}\right) \\
&= k(\theta)\cos\theta\sin\theta + \frac{(r' - r_0)\cos^2\theta}{r'} \cdot \frac{dk}{d\theta} + \frac{r\sin\theta\cos\theta}{r'}
\end{aligned}
\tag{S6}
$$

$$
\begin{aligned}
\frac{\partial y}{\partial x'} &= \frac{\partial y}{\partial r}\frac{\partial r}{\partial r'}\frac{\partial r'}{\partial x'} + \frac{\partial y}{\partial r}\frac{\partial r}{\partial \theta}\frac{\partial \theta}{\partial x'} + \frac{\partial y}{\partial \theta}\frac{\partial \theta}{\partial x'} \\
&= \sin\theta \cdot k(\theta)\cos\theta + \sin\theta \cdot (r' - r_0)\frac{dk}{d\theta} \cdot \left(-\frac{\sin\theta}{r'}\right) + (-r\cos\theta) \cdot \left(-\frac{\sin\theta}{r'}\right) \\
&= k(\theta)\sin\theta\cos\theta - \frac{(r' - r_0)\sin^2\theta}{r'} \cdot \frac{dk}{d\theta} - \frac{r\cos\theta\sin\theta}{r'}
\end{aligned}
\tag{S7}
$$

In the limit of small stretching $\varepsilon$, the stretching coefficient $k(\theta)$ is expanded according to Eq. S3:

$$
k(\theta) = \frac{(1+\varepsilon)R'(\theta) - r_0}{R'(\theta) - r_0} = 1 + \varepsilon\frac{R'(\theta)}{R'(\theta) - r_0} = 1 + \varepsilon C(\theta)
\tag{S8}
$$

where $C(\theta) = R'(\theta)/(R'(\theta) - r_0)$ is a bounded function independent of $\theta$ (for the square outer boundary, $C(\theta)$ ranges from 1 to 1.414). Differentiating with respect to $\theta$ yields the tangential transformation rate:

$$
\frac{dk}{d\theta} = \varepsilon \cdot \frac{dC}{d\theta}
\tag{S9}
$$

It is evident from the above expression that the tangential transformation rate is itself a first-order quantity in $\varepsilon$, which is the fundamental reason for the extremely low anisotropy.

We define two deviation quantities from the Cauchy-Riemann equations:

$$\Delta_1 = \left|\frac{\partial x}{\partial x'} - \frac{\partial y}{\partial y'}\right|, \quad \Delta_2 = \left|\frac{\partial x}{\partial y'} + \frac{\partial y}{\partial x'}\right| \tag{S10}$$

When $\Delta_1=0$ and $\Delta_2=0$, the transformation reduces to a strict conformal mapping.

1. The zeroth-order term ($\varepsilon=0$) corresponds to the identity transformation. When $\varepsilon=0$, we have $k(\theta)=1$ and $r=r'$. Substituting these into the partial derivative expression yields:

$$\begin{aligned}&\frac{\partial x}{\partial x'} = \cos^2\theta + \sin^2\theta = 1, \frac{\partial y}{\partial y'} = \cos^2\theta + \sin^2\theta = 1\\&\frac{\partial x}{\partial y'} = \cos\theta\sin\theta - \sin\theta\cos\theta = 0, \frac{\partial y}{\partial x'} = \sin\theta\cos\theta - \cos\theta\sin\theta = 0\end{aligned} \tag{S11}$$

Equation (S30) satisfies the Cauchy-Riemann equations exactly.

2. The first-order perturbation term（$\varepsilon<<1$）

Substituting $k(\theta)=1+\varepsilon C(\theta)$ and $dk/d\theta = \varepsilon\cdot dC/d\theta$ into the deviation expressions, all zeroth-order terms cancel exactly, leaving only the first-order terms in $\varepsilon$:

$$\begin{aligned}\Delta_1 &= \left|\frac{\varepsilon}{r'}\left[r_0 C(\theta)\left(\cos^2\theta - \sin^2\theta\right) - 2\left(r'-r_0\right)\cos\theta\sin\theta\cdot\frac{dC}{d\theta}\right]\right|\\ \Delta_2 &= \left|\frac{\varepsilon}{r'}\left[2r_0 C(\theta)\cos\theta\sin\theta + \left(r'-r_0\right)\left(\cos^2\theta - \sin^2\theta\right)\cdot\frac{dC}{d\theta}\right]\right|\end{aligned} \tag{S12}$$

It is evident from the above that each term in $\Delta_1=0$ and $\Delta_2=0$ contains the common factor $\varepsilon$, while the remaining coefficients $C(\theta)$, $dC/d\theta$, and the trigonometric functions are all finite and bounded. The common factor $\varepsilon$ in every term of $\Delta_1=0$ and $\Delta_2=0$ renders them infinitesimals of the same order.

3. Numerical verification (10% stretching, $\varepsilon=0.1$)

Substituting the typical parameters, the half-side length of the square in virtual space is set to $p_1/2=1$ (normalized, which does not affect the results), yielding $R'(\theta)=1/\cos\theta$. The central circle radius is taken as $r_0=0.2$ (i.e., the central circle diameter is 20% of the outer boundary, a conventional design for optical devices). Based on these, we obtain:

$$C(\theta)=\frac{1}{1-0.2\cos\theta}, \quad \frac{dC}{d\theta} = \frac{0.2\sin\theta}{\left(1-0.2\cos\theta\right)^2} \tag{S13}$$

The maximum deviation occurs at the four corner points in the positive directions ($\theta=45^\circ$). Substituting $\cos 45^\circ = \sin 45^\circ \approx 0.707$ into Eq. S13, we obtain:

$$C\left(45^\circ\right)=1.165, \quad \left.\frac{dC}{d\theta}\right|_{45^\circ} = 0.192 \tag{S14}$$

Substituting into the deviation expressions, we calculate:

$$\Delta_1 \approx 0.0330, \quad \Delta_2 \approx 0.0165 \tag{S15}$$

Final numerical conclusion: Under 10% uniaxial stretch, the maximum deviation from the Cauchy–Riemann equations is approximately 0.04, which matches the material's maximum anisotropy exactly.

Final conclusion (the material can be treated as approximately isotropic): For a radial linear scaling transformation with a fixed inner boundary and a slightly stretched outer boundary, as the stretch $\varepsilon \to 0$, the transformation rigorously reduces to a conformal mapping. When $\varepsilon \leq 0.1$ (10%), the maximum deviation from the Cauchy–Riemann equations remain within 4%, and the peak material anisotropy is likewise bounded by 4%. From an engineering perspective, the transformation can thus be reasonably approximated as conformal, permitting direct design using an isotropic equivalent permittivity.

Thus, we have demonstrated that under perturbative conditions, the aforementioned transformation can be regarded as approximately conformal.

**Second**, derive the expression for the dielectric tensor under conformal transformations.

In the transformation process of Fig. S1, the following conditions are satisfied: (1) There is no stretching/compression along the $z$ direction, i.e., $z = z'$. (2) The transformation is a smooth conformal transformation (angle-preserving). (3) The virtual space is vacuum. The Jacobian matrix is defined as the matrix of partial derivatives of the physical space coordinates with respect to the virtual space coordinates:

$$J = \frac{\partial X}{\partial X'} = \begin{bmatrix} \frac{\partial x}{\partial x'} & \frac{\partial x}{\partial y'} & \frac{\partial x}{\partial z'} \\ \frac{\partial y}{\partial x'} & \frac{\partial y}{\partial y'} & \frac{\partial y}{\partial z'} \\ \frac{\partial z}{\partial x'} & \frac{\partial z}{\partial y'} & \frac{\partial z}{\partial z'} \end{bmatrix} \tag{S16}$$

Since there is no stretching in the $z$ direction, i.e., $z = z'$, it directly follows that all partial derivatives with respect to $z$ are zero:

$$\frac{\partial x}{\partial z'} = 0, \frac{\partial y}{\partial z'} = 0, \frac{\partial z}{\partial x'} = 0, \frac{\partial z}{\partial y'} = 0, \frac{\partial z}{\partial z'} = 1 \tag{S17}$$

Therefore, the Jacobian matrix reduces to a block diagonal matrix:

$$J = \begin{pmatrix} A & 0_{2\times1} \\ 0_{1\times2} & 1 \end{pmatrix} \tag{S18}$$

where $A = \partial(x, y)/\partial(x', y')$. From $\varepsilon = \varepsilon_0 JJ^T/\det(J)$, the full expression of the dielectric tensor is derived as:

$$\varepsilon = \frac{1}{\det(J)} \begin{pmatrix} \left(\frac{\partial x}{\partial x'}\right)^2 + \left(\frac{\partial x}{\partial y'}\right)^2 & \frac{\partial x}{\partial x'}\frac{\partial y}{\partial x'} + \frac{\partial x}{\partial y'}\frac{\partial y}{\partial y'} & 0 \\ \frac{\partial y}{\partial x'}\frac{\partial x}{\partial x'} + \frac{\partial y}{\partial y'}\frac{\partial x}{\partial y'} & \left(\frac{\partial x}{\partial x'}\right)^2 + \left(\frac{\partial x}{\partial y'}\right)^2 & 0 \\ 0 & 0 & 1 \end{pmatrix} \tag{S19}$$

It is well known that the necessary and sufficient condition for a conformal transformation is that it satisfies the Cauchy-Riemann equations. The Cauchy-Riemann equations are:

$$\frac{\partial x}{\partial x'}=\frac{\partial y}{\partial y'},\frac{\partial x}{\partial y'}=-\frac{\partial y}{\partial x'} \tag{S20}$$

Substituting the Cauchy-Riemann conditions into the Jacobian determinant:

$$\det(A)=\frac{\partial x}{\partial x'}\frac{\partial y}{\partial y'}-\frac{\partial x}{\partial y'}\frac{\partial y}{\partial x'}=\left(\frac{\partial x}{\partial x'}\right)^2+\left(\frac{\partial x}{\partial y'}\right)^2 \tag{S21}$$

Similarly, we obtain:

$$\det(A)=\left(\frac{\partial y}{\partial x'}\right)^2+\left(\frac{\partial y}{\partial y'}\right)^2 \tag{S22}$$

Calculating the off-diagonal components:

$$\frac{\partial x}{\partial x'}\frac{\partial y}{\partial x'}+\frac{\partial x}{\partial y'}\frac{\partial y}{\partial y'}=\frac{\partial x}{\partial x'}\left(-\frac{\partial x}{\partial y'}\right)+\frac{\partial x}{\partial y'}\cdot\frac{\partial x}{\partial x'}=0 \tag{S23}$$

The dielectric tensor is finally simplified to:

$$\varepsilon=\frac{\varepsilon_0}{\det(J)}=\begin{pmatrix}\det(J) & 0 & 0\\ 0 & \det(J) & 0\\ 0 & 0 & 1\end{pmatrix}=\begin{pmatrix}\varepsilon_0 & 0 & 0\\ 0 & \varepsilon_0 & 0\\ 0 & 0 & \dfrac{\varepsilon_0}{\det(J)}\end{pmatrix} \tag{S24}$$

From Eq. S24, it can be seen that under a conformal transformation, the dielectric constant tensor automatically reduces to a diagonal tensor.

By the same reasoning, the expression for the permeability tensor is:

$$\mu=\frac{\mu_0}{\det(J)}=\begin{pmatrix}\det(J) & 0 & 0\\ 0 & \det(J) & 0\\ 0 & 0 & 1\end{pmatrix}=\begin{pmatrix}\mu_0 & 0 & 0\\ 0 & \mu_0 & 0\\ 0 & 0 & \dfrac{\varepsilon_0}{\det(J)}\end{pmatrix} \tag{S25}$$

**Third**, from the divergence equations of the electromagnetic fields, it is derived that a TEM electromagnetic field normally incident on a two-dimensional metasurface must generate polarization in the $E_z$ direction.

The total field in region ($z<0$) is the superposition of the incident field and the reflected field. Since the metasurface is a periodic structure, the reflected field can be expanded into a Fourier series (Bloch wave expansion):

$$\vec{E}_1(x,y,z)=\vec{E}^{inc}+\sum_{m,n=-\infty}^{\infty}\vec{E}_{mn}^{ref}e^{i\left(k_{xmn}x+k_{ymn}y-k_{zmn}z\right)} \tag{S26}$$

where the transverse component of the Bloch vector is:

$$k_{xmn}=\frac{2\pi m}{p},k_{ymn}=\frac{2\pi n}{p} \tag{S27}$$

Where $m$, $n$ are the diffraction orders, where $m=n=0$ corresponds to the zeroth-order reflected wave (specular reflection), and $|m|$, $|n|\geq 1$ correspond to higher-order diffracted waves. The total field inside the metasuface ($0<z<d$) can also be expanded as a Bloch series:

$$\vec{E}_2(x,y,z)=\sum_{m,n=-\infty}^{\infty}\vec{E}_{mn}^{\text{int}}e^{i\left(k_{xmn}x+k_{ymn}y+k'_{zmn}z\right)} \tag{S28}$$

The transverse wave vectors are exactly the same as those in region ($z<0$), which is a strict requirement of the periodic boundary conditions. At $z=0$, substituting the tangential electric field continuity into the expansion on both sides:

$$E_0+\sum_{m,n}E_{x,mn}^{ref}e^{i\left(k_{xmn}x+k_{ymn}y\right)}=\sum_{m,n}E_{x,mn}^{\text{int}}e^{i\left(k_{xmn}x+k_{ymn}y\right)} \tag{S29}$$

Since $\varepsilon_{zz}(x,y)$ is not a constant, the metasurface scatters high-order diffracted waves. Thus, there exists at least one set of $(m,n)\neq(0,0)$ such that $E_{x,mn}^{\text{int}}\neq 0$ is satisfied.

Since the transverse electric field $E_x$ $(x,y,z)$ inside the metasurface is a periodic function of $x$ and $y$ containing higher-order harmonic components, equations $\partial E_x/\partial x\neq 0$ and $\partial E_y/\partial x\neq 0$ are thus satisfied. Inside the metasurface ($0<z<d$), the source-free divergence equation $\nabla\cdot\vec{D}=0$ holds:

$$\frac{\partial D_x}{\partial x}+\frac{\partial D_y}{\partial y}+\frac{\partial D_z}{\partial z}=0 \tag{S30}$$

Substituting $D_x=\varepsilon_0E_x$, $D_y=\varepsilon_0E_y$, $D_z=\varepsilon_0E_z$, and canceling $\varepsilon_0$:

$$\frac{\partial E_x}{\partial x}+\frac{\partial E_y}{\partial y}+\frac{\partial}{\partial z}\left(\varepsilon_{zz}E_z\right)=0 \tag{S31}$$

We have already proven that $\partial E_x/\partial x+\partial E_y/\partial x\neq 0$ holds (due to the existence of higher-order harmonics). Thus, the first term on the left-hand side of the source-free divergence Eq. S31 is non-zero. To satisfy the divergence equation, the second term must also be non-zero:

$$\frac{\partial}{\partial z}\left(\varepsilon_{zz}E_z\right)\neq 0 \tag{S32}$$

From the above Eq. S32, it follows that:

$$E_z(x,y,z)\neq 0,0<z<d \tag{S33}$$

In summary, we have proven that inside the metasurface, $E_z\neq 0$. The entire derivation involves no assumptions and follows entirely from the boundary conditions and Maxwell's equations.

As shown in Eq. S24 and Eq. S25, the dielectric tensor ε and permeability tensor μ is a diagonal tensor. The components of dielectric tensor ($\varepsilon_{xx}=\varepsilon_{yy}=\varepsilon_0$) and components of *permeability tensor* ($\mu_{xx}=\mu_{yy}=\mu_0$) in the $x$ and $y$ directions are equal to $\varepsilon_0$ and $\mu_0$, *respectively*. Therefore, the transverse waves in the $x$ and $y$ directions within the metasurface are perfectly matched with the vacuum impedance, presenting as pure traveling waves with no reflection, and will not form standing waves between the interface. The essence of resonance is the accumulation of the energy of the standing wave. Therefore, no resonance occurs in the transverse fields. According to Eq. (24) and Eq. (25), there is a phase mismatch in the longitudinal field, so the longitudinal field $E_z$ can produce resonance, and we have already proved in Eq. S33 that the longitudinal field is non-zero. Given that the resonant component is mainly focused, the tensor in Eq. S24 can be reduced to a scalar, that is:

$$\varepsilon_{zz} = \frac{1}{\det(J)} \tag{S34}$$

The definition of the equivalent dielectric constant is the area-weighted average, and the expression of the equivalent dielectric constant $\varepsilon_{\mathrm{eq}}$ is:

$$\varepsilon_{eq} = \frac{1}{S_p}\iint_{S_p} \varepsilon_{zz} \cdot dS \tag{S35}$$

$S_p$ and $S_v$ denote the air-gap areas in Figs. 2(f) and 2(e), defined as physical space and virtual space with area elements $dS$ and $dS'$, respectively. By calculus, $dS$=det($J$)$dS'$. Substituting into Eq. S34–S34 yields:

$$\iint_{S_p} \varepsilon_{zz} \cdot dS = \varepsilon_0 \iint_{S_v} \frac{1}{\det(J)} \det(J)\, dS' = \varepsilon_0 \cdot S_v \tag{S36}$$

By combining Eq. S35 and Eq. S36, we can obtain:

$$\varepsilon_{eq} = \varepsilon_0 \frac{S_v}{S_p} = \varepsilon_0 \frac{p_2^{\ 2} - S_{\mathrm{unit}}}{p_1^{\ 2} - S_{unit}} \tag{S37}$$

Eq. S37 corresponds to Eq. (1) in the main text. Thus, Eq. (1) is obtained.

## S2: Deformation of the metasurface gap along a single direction

Figure S2 illustrates the physical picture under of the deformation of metasurface gap along a single direction. As shown in Figs. S2(d)and S2 (f), representative unit cells are extracted from metasurfaces A and B, with lengths $p_1$ and $p_2$, respectively, and identical widths $p_1$. The cylindrical meta-atoms share a uniform diameter. Thus, the two configurations differ only in their gap dimensions along one direction.

The physical picture revealed by the mechanism is shown in Fig. S2, which involves transforming metasurface B into B′ by stretching its gaps to match the period of metasurface A. Based on transformation optics, B and B′ are electromagnetically equivalent, allowing the study of the A−B relationship to be reformulated as the A−B′ relationship. Since A and B′ share identical periodicities, the only difference is the dielectric tensor of the air gap due to space stretching. For small variations, this permittivity change is treated as a perturbation. Thus, perturbation theory using metasurface A as a reference yields the resonant frequency of metasurface B′, which equals that of metasurface B due to their electromagnetic equivalence. A Cartesian coordinate system is defined as follows: the electromagnetic wave is normally incident along the $z$-axis, with the metasurface lying in the $x$-$y$ plane.

Similarly to the derivations in grating, Thus, Eq. S38 can be obtained according to Fig. S2:

$$\frac{\Delta f}{f_0} = -\frac{\varepsilon_0 \varepsilon_r}{2W}\int_{\Delta V_{\mathrm{unit}}} \Delta\varepsilon_r \left|\vec{E}_0\right|^2 dV \approx \begin{cases} c_x(1-\beta), x - polarization \\ c_y(1-1/\beta) + c_z(1-\beta), y - polarizaiton \end{cases} \tag{S38}$$

where $f_0$ and $f$ denote the resonant frequencies at metasurface periods of $p_1$ and $p_2$, respectively. $\Delta f = f - f_0$ denotes the shift in resonant frequency after the gap is distorted. $\Delta V_{\mathrm{gap}}$ is the volume of gap in metasurface, $\beta = p_2/p_1$ is the scaling factor. As shown in Fig. S2, the metasurface is uniaxially compressed along the $y$-direction (during the compression, the unit size the meta-atom remains unchanged, with only the air gap being compressed). $W$ denotes the total electromagnetic energy distributed over one period of the metasurface. When the change of the gap space involved in the above transformation is sufficiently small to be treated as a perturbation, the ratio of the energy stored in the air gap to the total energy $W$ within one period of the metasurface is approximately constant. Thus, the $c_i = \varepsilon_0 \iint_{\Delta V gap} \left|\vec{E}_{0i}\right|^2 dV \Big/ 2W \; (i = x, y, z)$ can be seen constants. Note that $c_i$ an unknown constant. By obtaining the resonant frequencies of three metasurfaces with different periods via full-wave simulation, $c_i$ can be solved from Eq. (S38).

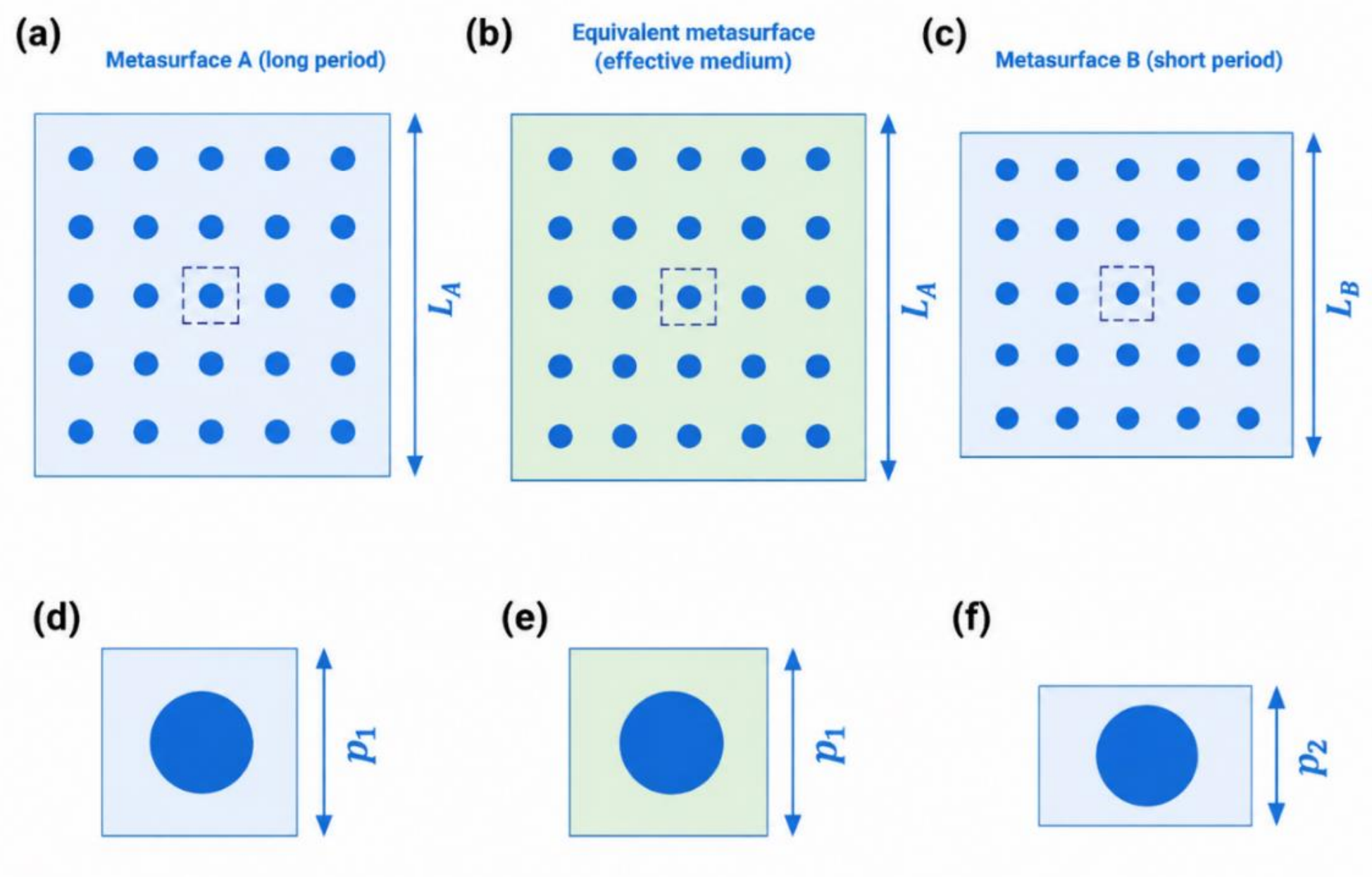

**Fig. S2.** Physical picture of deformation of the metasurface gap along $y$-axis: stretching the gap from metasurface B to B′ induces spatial deformation, which causes permittivity perturbation. Consequently, metasurface B′ can be treated as a perturbation on metasurface A.

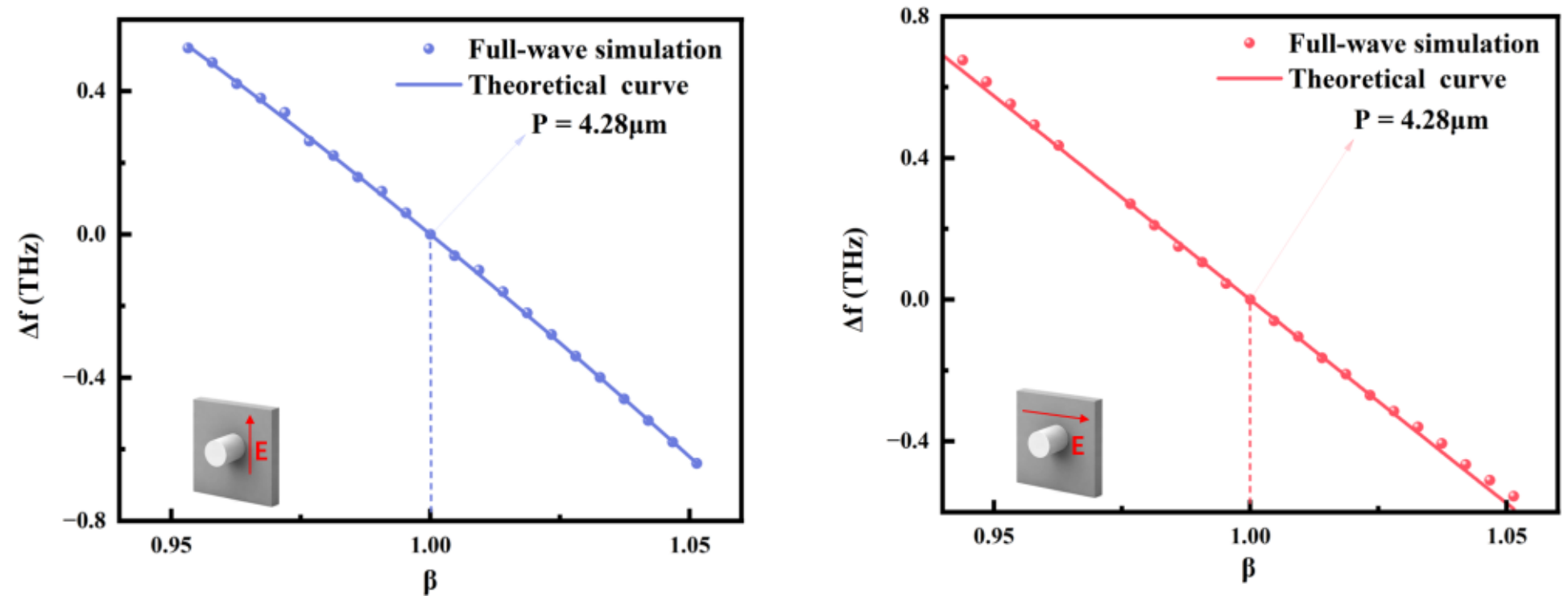

**Fig. S3.** (a) and (b) are the theoretical curve and the full-wave simulation result for $y$- and $x$-polarized incidences under deformation of the metasurface gap along a single direction.

Figure S3 shows that keeping meta-atom size fixed while deforming the metasurface gap along a single direction (thus compressing the air gap). It is verified for a metasurface with cylindrical meta-atoms ($r$ = 1 μm, $h$ = 3 μm). And the period of the metasurface is 4.28 μm. Figure S3 further reveals that uniaxial compression (with scaling factor β along $x$-axis) induces anisotropic responses to different polarizations. Specifically, Figs. S3a and S3b show the response curves for $y$- and $x$-polarized normal incidence, respectively, which compares full-wave simulations (data points) with theoretical predictions (curves), showing good agreement. These results further validate the mechanism in deformation of the meta- gap along a single direction.

## S3: Deformation of the meta-atom along a single direction

Figure S4 illustrates the physical picture under of the deformation of meta-atom along a single direction. As shown in Figs. S4(d)-S4(f), representative unit cells are extracted from metasurfaces A, B and B′ with identical side length $p_1$. A Cartesian coordinate system is defined as follows: the electromagnetic wave is normally incident along the $z$-axis, with the metasurface lying in the $x$-$y$ plane.

The physical picture revealed by the mechanism is shown in Fig. S4, which involves transforming metasurface B into B′ by stretching the meta-atom to match the meta-atom of metasurface A. Figs.S4(d)-S4(f) are one unit cell extracted from Figs. S4(a)-4(c). The period is fixed at $p_1$, with unit widths $h_1$ (Fig. S4(d)) and $h_2$ (Fig. S4(f)) for metasurfaces A and B, respectively. The width difference makes direct correlation between metasufrace A and B nontrivial. This mechanism resolves this by transforming metasurface B′ to match the unit cell of metasurface A. Via transformation optics, metasurface B′ is electromagnetically equivalent to B, reducing the problem to comparing A with B′. As seen in Figs. S4(d)–S4(e), the only distinction is a slight permittivity variation in the meta-atom. Perturbation theory can then quantify its impact.

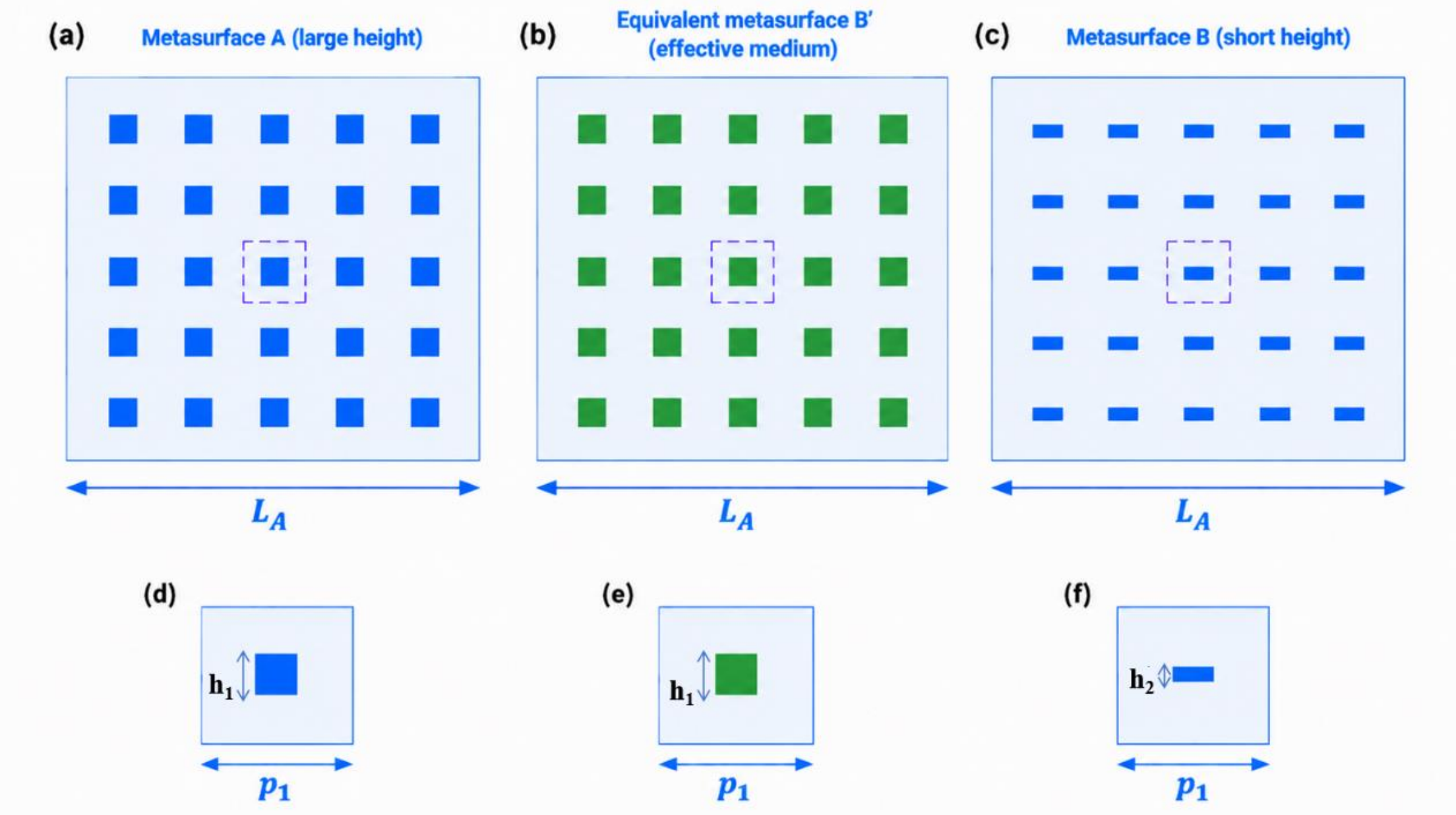

**Fig. S4.** Physical picture of deformation of the meta-atom along $y$-axis: stretching the meta-atom from metasurface B to B′ induces spatial deformation, which causes permittivity perturbation. Consequently, metasurface B′ can be treated as a perturbation on metasurface A.

Similarly to the derivations in grating, Eq. S39 can be obtained:

$$\frac{\Delta f}{f_0} = -\frac{\varepsilon_0 \varepsilon_r}{2W} \int_{\Delta V_{\text{unit}}} \Delta \varepsilon_r \left| \vec{E}_0 \right|^2 dV \approx \begin{cases} c_x (1-\beta), x - polarization \\ c_y (1-1/\beta) + c_z (1-\beta), y - polarizaiton \end{cases} \tag{S39}$$

where $f_0$ and $f$ denote the resonant frequencies at meta-atom of $h_1$ and $h_2$, respectively. $\Delta f = f - f_0$ denotes the shift in resonant frequency after the gap is distorted. $\Delta V_{\text{unit}}$ is the volume of meta-atom, $\beta = h_2/h_1$ is the scaling factor of the meta-atom. As shown in Fig. S4, the metasurface is uniaxially compressed along the *y*-direction (during the compression, the air gap remains unchanged, with only the width of meta-atom is compressed along the *y*-axis). Similarly, the $c_i = \varepsilon_0 \varepsilon_r \iint_{\Delta V unit} \left| \vec{E}_{0i} \right|^2 dV \Big/ 2W \ (i = x, y, z)$ can be seen constants. Note that $c_i$ an unknown constant. By obtaining the resonant frequencies of three metasurfaces with different periods via full-wave simulation, $c_i$ can be solved from Eq. S39.

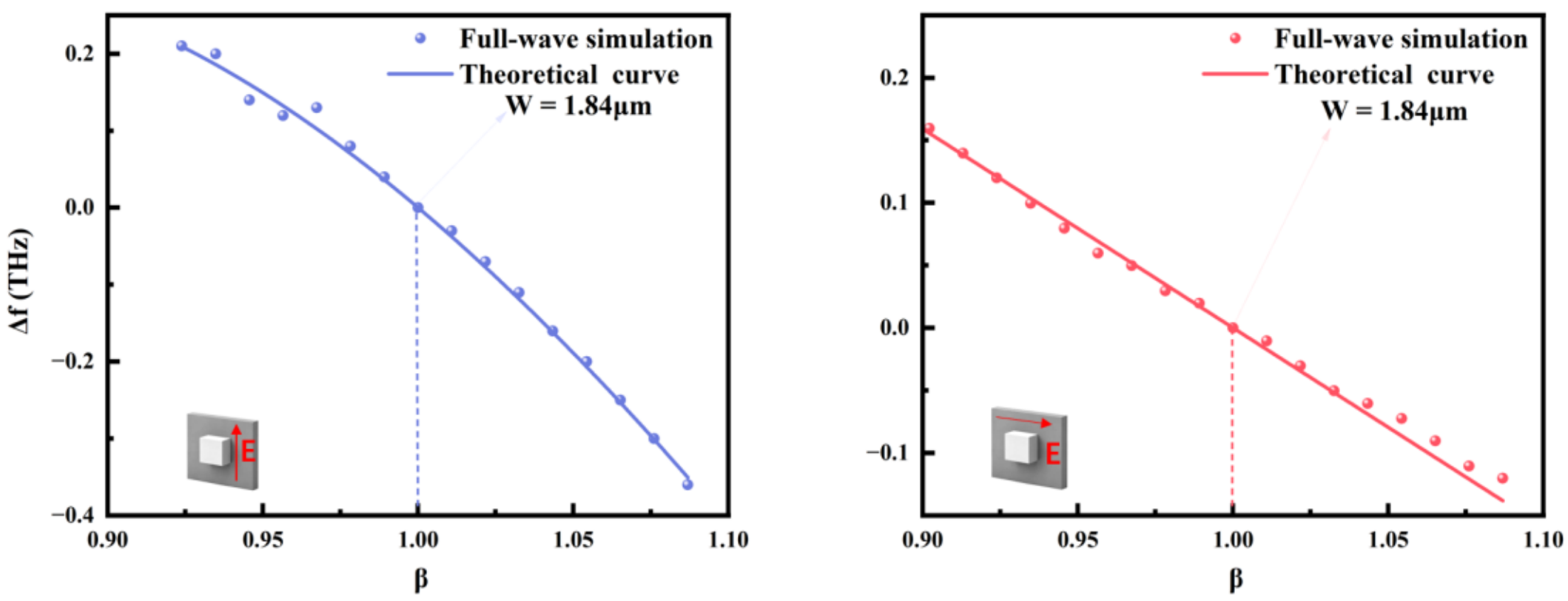

**Fig. S5.** (a) and (b) are the theoretical curve and the full-wave simulation result for *y*- and *x*-polarized incidences under deformation of the meta-atom along a single direction.

Figure S5 shows that the period of the metasurface fixed while deforming the meta-atom along a single direction. It is verified for a metasurface with square meta-atoms ($w$= 2 μm). Figure S5 further reveals that uniaxial compression (with scaling factor β along *y*-axis) induces anisotropic responses to different polarizations. Specifically, Figs. S5(a) and S5(b) show the response curves for *y*- and *x*-polarized normal incidence, respectively, which compares full-wave simulations (data points) with theoretical predictions (curves), showing good agreement. These results further validate the mechanism in deformation of the meta-atom along a single direction.